\documentclass[12pt]{article}

\usepackage[letterpaper,hmargin=1in,vmargin=1in]{geometry}

\usepackage{graphicx,epstopdf,amsmath,amsfonts}

\def\be{\begin{equation}}
\def\ee{\end{equation}}
\def\ba{\begin{eqnarray}}
\def\ea{\end{eqnarray}}
\def\ge{\mathrel{\raise.3ex\hbox{$>$\kern-.75em\lower1ex\hbox{$\sim$}}}}
\def\la{\mathrel{\raise.3ex\hbox{$<$\kern-.75em\lower1ex\hbox{$\sim$}}}}

\def\simgt{\mathrel{\raise.3ex\hbox{$>$\kern-.75em\lower1ex\hbox{$\sim$}}}}
\def\simlt{\mathrel{\raise.3ex\hbox{$<$\kern-.75em\lower1ex\hbox{$\sim$}}}}

\newcommand{\bi}[1]{\bibitem{#1}}
\newcommand{\fr}[2]{\frac{#1}{#2}}

\newcommand{\nc}{\newcommand}

\nc{\gone}{\bar g_{\pi NN}^{(1)}}
\nc{\gzero}{\bar g_{\pi NN}^{(0)}}
\nc{\al}{\alpha}
\nc{\ga}{\gamma}
\nc{\de}{\delta}
\nc{\ep}{\epsilon}
\nc{\ze}{\zeta}
\nc{\et}{\eta}
\nc{\Th}{\Theta}
\nc{\ka}{\kappa}
\nc{\rh}{\rho}
\nc{\si}{\sigma}
\nc{\ta}{\tau}
\nc{\up}{\upsilon}
\nc{\ph}{\phi}
\nc{\ch}{\chi}
\nc{\ps}{\psi}
\nc{\om}{\omega}
\nc{\Ga}{\Gamma}
\nc{\De}{\Delta}
\nc{\La}{\Lambda}
\nc{\Si}{\Sigma}
\nc{\Up}{\Upsilon}
\nc{\Ph}{\Phi}
\nc{\Ps}{\Psi}
\nc{\Om}{\Omega}
\nc{\ptl}{\partial}
\nc{\del}{\nabla}
\nc{\ov}{\overline}
\nc{\newcaption}[1]{\centerline{\parbox{15cm}{\caption{#1}}}}

\nc{\hef}{$^4$He}
\nc{\het}{$^3$He}
\nc{\lisx}{$^6$Li}
\nc{\lisv}{$^7$Li}
\nc{\bes}{$^7$Be}
\nc{\beet}{$^8$Be}
\nc{\bor}{$^9$B}
\nc{\ben}{$^9$Be}
\nc{\ctw}{$^{12}$C}
\nc{\hefm}{{\rm ^4He}}
\nc{\hetm}{{\rm ^3He}}
\nc{\lisxm}{{\rm ^6Li}}
\nc{\lisvm}{{\rm ^7Li}}
\nc{\besm}{{\rm ^7Be}}
\nc{\beetm}{{\rm ^8Be}}
\nc{\borm}{{\rm ^9B}}
\nc{\benm}{{\rm ^9Be}}
\nc{\bs}{(N$X^-$)}
\nc{\xm}{$X^-$}
\nc{\xp}{$X^+$}
\nc{\xz}{$X^0$}
\nc{\bex}{(\bes\xm)}
\nc{\bexm}{(\besm X^-)}

\begin{document}

\begin{titlepage}

\setcounter{page}{1}

\vspace*{0.2in}

\begin{center}

\hspace*{-0.6cm}\parbox{17.5cm}{\Large \bf Resonant 
enhancement of nuclear reactions as a possible
solution to the cosmological lithium problem }

\end{center}

\vspace*{0.5cm}
\normalsize

\begin{center}

{\bf Richard H. Cyburt$^{\,(a,b)}$}  and {\bf Maxim Pospelov$^{\,(c,d)}$}

\smallskip
\medskip

$^{\,(a)}${\it Joint Institute for Nuclear Astrophysics (JINA) {\tt http://www.jinaweb.org}}

$^{\,(b)}${\it National Superconducting Cyclotron Laboratory, Michigan State University, \\
     East Lansing, MI, 48824 U.S.A.}

$^{\,(c)}${\it Department of Physics and Astronomy, University of Victoria, \\
     Victoria, BC, V8P 1A1 Canada}

$^{\,(d)}${\it Perimeter Institute for Theoretical Physics, Waterloo,
ON, N2J 2W9, Canada}

\smallskip
\end{center}
\vskip0.2in

\centerline{\large\bf Abstract}
There is a significant discrepancy between the current theoretical
prediction of the cosmological lithium abundance, mostly produced as
\bes\ during the Big Bang, and its observationally inferred value. We
investigate whether the resonant enhancement of \bes\ burning
reactions may alleviate this discrepancy. We identify one narrow
nuclear level in $^9$B, $E_{5/2^+} \simeq 16.7$ MeV that is not
sufficiently studied experimentally, and being just $\sim 200$ keV
above the \bes+d threshold, may lead to the resonant enhancement of
\bes$(d,\gamma$)\bor\ and
\bes$(d,p)\alpha\alpha$ reactions. We determine 
the relationship between the domain of resonant energies $E_r$ and the
deuterium separation width $\Gamma_d$ that results in the significant
depletion of the cosmological lithium abundance and find that $ (E_r,
~\Gamma_{d}) \simeq (170-220,~10-40)$ keV can eliminate current
discrepancy. Such a large width at this resonant energy can be only
achieved if the interaction radius for the deuterium entrance channel
is very large, $a_{27} \ge 9$ fm. Our results also imply that before
dedicated nuclear experimental and theoretical work is done to
clarify the role played by this resonance, the current conservative
BBN prediction of lithium abundance should carry significantly larger
error bars, [\lisv/H]$_{\rm BBN}$$= (2.5-6)\times 10^{-10}$.

\vfil
\leftline{June 2009}

\end{titlepage}

\section{ Introduction}
The enhancement of nuclear reaction rates by nuclear resonances is
extremely important in nuclear astrophysics, as was first brilliantly
demonstrated by Hoyle in the early 1950-s. The remarkable prediction
of the 7.65 MeV resonance in $^{12}$C not far from the \beet\ + \hef\
separation threshold was based on the consideration of the carbon
abundance. The subsequent experimental discovery of this level was one
of the defining moments in nuclear astrophysics, and catalyzed further
experimental and theoretical work in that area of science
\cite{Fowler}.  More than half a century after this memorable chapter
in the history of physics, it is tempting to speculate whether "history can repeat
itself". In this paper we consider a possibility that resonant
enhancements of nuclear reaction rates could be responsible for the
solution of the so-called cosmological lithium problem.

The problem with the \lisv\ abundance came to light during the last
decade, as the fast experimental progress in cosmology, and in
particular with detailed studies of the anisotropies in the cosmic
microwave background (CMB) allowed to sharpen the determination of
many cosmological parameters. Currently, among one of the best known
parameters is the ratio of the baryon and photon number densities,
$\eta_b^{WMAP} = (6.23 \pm 0.17)\times 10^{-10}$ \cite{WMAP}.  Since
$\eta_b$ is the main cosmological input into the standard theory of
primordial nucleosynthesis (BBN) \cite{Sarkar}, the uncertainties in
predictions of primoridal helium, deuterium and lithium abundances
shrank to unprecedented levels. In particular, the quality of the
comparison of the predicted helium mass fraction $Y_p$ and primordial
deuterium abundance with observations mostly depends on the erorrs in
extracting the primordial fractions for these elements from
observational data. Currently, there is no disagreement between the
predicted and observationally extracted primordial abundances of these
two elements.

In conrast to the helium and deuterium abundances, predicted and
observed lithium abundances exhibit a serious discrepancy. Its
primordial fraction is inferred from the observations of the lithium
absorption lines in the atmospheres of population II stars, where as
function of metallicity (at low metallicity), lithium exhibits
remarkable constancy, known as the Spite plateau
\cite{SpSp}. Extrapolation of the Spite plateau to zero metallicity is
believed to reflect the primordial value of lithium. Current status of
the lithium problem can be summed as follows:
\begin{eqnarray}
\label{Spite}
{\rm Spite~plateau~value:}&&~~\fr{\lisvm}{\rm H} = 1.23^{+0.34}_{-0.16}\times 10^{-10}
\\
{\rm BBN~theory}:&&~~\fr{\lisvm}{\rm H} = 5.24^{+0.71}_{-0.67}\times 10^{-10},
\label{BBNvalue}
\end{eqnarray}
where we use the results of evaluation of lithium abundance in field stars 
\cite{Ryan} and the latest theoretical BBN evaluation \cite{Cyburt:2008}, 
both shown as 68\% confidence limits (\cite{Ryan} 95\% error reduced
by half for 68\% confidence). It is important to keep in mind that
measurements of lithium in globular clusters have resulted in somewhat
higher abundances $(2.19\pm 0.28)\times 10^{-10}$
\cite{Bonifacio} (for other observational determinations of \lisv\
abundance consistent with \cite{Ryan,Bonifacio} see
\cite{otherworks,Asplund}). Although there were some claims that the
re-calibration of the effective temperature scale is needed that leads
to a $\sim$50\% increase in the resulting lithium abundance \cite{MR},
later studies \cite{MRfollowup} do not find support to this
suggestion.

Another interesting twist to the lithium story is added by the claim
of the detection of the \lisx\ metallicity plateau at \lisx/H $\sim
O(10^{-11})$ level, which almost certainly implies some form of
pre-galactic \lisx. The significant presence of \lisx\ would also have
serious implications for any stellar mechanism that is able to deplete
\lisv, as \lisx\ is more fragile and is destroyed at lower
temperatures. The status of \lisx\ plateau claim so far remains in
doubt, as subsequent more conservative analyses found no evidence for
the plateau \cite{antiAsplund}. At this point, it is fair to say that
only \lisv\ presents a serious conflict between standard cosmology and
observations.

There are several logical possibilities of how the \lisv\ discrepancy
can be resolved that are actively discussed in the literature. It is
plausible that the resolution of the lithium problem could involve
astrophysical effects, nuclear physics effects or completely new
effects in particle physics or cosmology, or any combination of the
above. Below we outline some work that has been done in these
directions:

\begin{itemize}

\item {\em Astrophysical resolution.} It is possible that the cosmological abundance of lithium 
is altered in the subsequent evolution. Most notably, there were
suggestions that population II stars themselves deplete lithium
\cite{Korn} through {\em e.g.} diffusional settling during their main sequence 
lifetimes. Although such a possibility cannot be excluded, any
hypothetical stellar process that depletes lithium by a factor of 2-3
should preserve low scatter along the Spite plateau, and be almost
independent on other varying astrophysical parameters such as amount
of stellar rotation, variations in temperature, etc.

\item {\em Nuclear physics resolution.} Most of primordial lithium is produced as \bes.
It is conceivable that some of the 
nuclear physics reactions that affect \bes\ abundance are mismeasured
or miscalculated. Recent analyses of all relevant reaction rates do
not support this possibility \cite{CFO,french,Cyburt,serpico}, as most of the
important reactions are measured/calculated by several groups. Only a
drastic change to some secondary reactions, {\em e.g.} the enhacement
of \bes($d,p$)$\alpha\alpha$ rate by a factor of $O(100)$ over the one
in the Caughlan-Fowler compilation, could deplete overall \bes\
abundance \cite{french}. The follow-up experiment to check this
hypothesis has produced a negative result \cite{Angulo}.

\item {\em New Physics resolution.} Less likely at this point, but the new physics 
resolution of the lithium problem may indeed be contemplated. One example
are the decays of heavy meta-stable particles that inject extra
neutrons at $T\sim 40$ keV, which enhances the destruction of \bes\
\cite{unstable}. Meta-stable negatively charged particles may also lead
to the reduction of lithium abundance through the catalysis of
reactions that destroy \bes\ \cite{CBBN}. Other, more radical ideas
include the variation of the strong interaction coupling in time along
with modification of the gravitational sector \cite{crazyideas}. If
indeed the resolution of the lithium problem is related to the
meta-stable charged weak-scale particles, there is some hope of testing
such hypothesis at the LHC.

\end{itemize}

We would like to emphasize that the search for the origin of \lisv\
discrepancy is extremely important for the consistency of modern
cosmology. It may hold a clue for the modification of the standard
cosmological framework, or at the very least lead to a new level of
understanding of physical processes in stellar atmospheres.

In this paper we revisit the possibility that nuclear physics is
responsible for the current discrepancy, and investigate whether the
primordial value for \lisv+\bes\ can be reduced. Unlike
Ref. \cite{french}, that introduced arbitrary rescaling factors to the
reaction rates, we take a different approach, having a closer look at
the resonances involved in many reactions. Giving credit to the 50
years of progress in nuclear astrophysics since Hoyle's discovery, we
are {\em not} going to hypothesize any new resonances, but explore the
possibilities that the existing identified resonances could lead to
the \bes\ depletion. In the next section, we list the resonances that
are known to affect \bes\ abundance or have some, perhaps remote,
possibility of affecting it. We then analyze whether the enhancement
of their strength is possible and find one candidate, the 16.7 MeV
resonance in \bor\, that can lead to the enhancement of the \bes\
burning in reaction with deuterium. In section 3, we modify the BBN
code to include this resonance, and determine the
parameters of the resonance that are required to achieve the
concordance of the BBN output with the Spite plateau value. We find
that indeed such adjustment is theoretically possible, albeit at the
very end of the reasonable range for the maximally allowed deuterium
separation widths. Section 4 contains the assessment of the
uncertainty of the BBN-predicted value for \lisv+\bes, given the
uncertainty in the property of this resonance. It also calls for
direct experimental determination of the properties of this resonance,
and the corresponding reaction rates. These future efforts might
either support our hypothesis thus offering a nuclear physics solution
to the lithium problem, or refute it, closing perhaps
the last nuclear "loophole" in the BBN prediction of lithium
abundance.

\section{Nuclear resonances affecting lithium abundance}

The 2004 re-analysis of nuclear rates \cite{Cyburt} have produced the
following scaling relation for the predicted total value of lithium
for the WMAP-I input value of $\eta_b$:
\ba \nonumber
10^{10}\fr{\lisvm}{\rm H} &=& 4.364\! \left(\! \fr{\eta_b}{6.14\!\times\! 10^{-10}}\!\right)^{2.12}\!\!
\left(\! \fr{\tau_n}{\tau_{n,0}}\!\right)^{0.44}\!\!\left(\! \fr{G_N}{G_{N,0}}\!\right)^{-0.72} \\
 &\times& r_2^{1.34}r_9^{0.96}r_8^{-0.76}r_{11}^{-0.71}r_4^{0.71}r_3^{0.59}r_6^{-0.27}.
\label{scaling}
\ea
Here $r_i$ are the reaction rates $R_i$ in the nomenclature of
Ref. \cite{Cyburt}, normalized on the 2004 recommended reaction
rates. Besides reaction rates and $\eta_b$ dependence, (\ref{scaling})
also contains the neutron lifetime and Newton's gravitational
constant, normalized to their measured values.

As it is well known, at the CMB-determined baron asymmetry most of lithium
is produced as \bes. The most important temperature range is $T_9
\simeq 0.3-0.6$ ($T_9\equiv T/10^9$K) where the main production and
destruction mechanisms are as follows:
\begin{eqnarray}
\label{SBBNprod}
{\rm Production}, R_9:&&~~~ \hetm(\alpha,\gamma)\besm
\\ 
{\rm Destruction}, R_{11}+R_{12} &&~~~ \besm(n,p)\lisvm;~~\lisvm(p,\alpha)\hefm.
\label{SBBNdest}
\end{eqnarray}

Recent progress in re-measuring \cite{He3He4} and re-analyzing
\cite{CD} $R_9$, the rate for reaction (\ref{SBBNprod}), allowed to
bring the total uncertainty in the production rate below the 10\%
level, and together with a revised determination of $\eta_b$ led to a
slight increase in the predicted lithium abundance,
Eq. (\ref{BBNvalue}). Other important parameters that regulate
total \bes\ abundance is the availability of free neutrons, the
abundance of \het\ at $T_9 \simeq 0.5 $, and the neutron capture rate
on \bes. The abundance of \het\ and availability of neutrons do depend
of course on the reaction rates among $A\leq 4$ elements, as
reflected in (\ref{scaling}). It is important to emphasize that all
the reaction rates in (\ref{scaling}) are known with better than 10\%
accuracy at BBN temperatures.

\begin{table}
\label{t1}
\begin{center}
\begin{tabular}{c|c|c|c}
Resonance & Reactions & $E_r$[keV] & $\Gamma_{cm}$[keV]\\
\hline\hline
 $np$, ground state of $^1S$   & $R_2:$ $n(p,\gamma)d$ & $\simeq 67$ &  $\sim 40$ \\  
 $^5$Li, $3/2^+$, 16.87 MeV & $R_8:$ $\hetm(d,p)\hefm$ & $\simeq 210$ &  $\simeq 270$  \\
$\beetm $, $2^-$, 18.91 MeV & $R_{11}:$  $ \besm(n,p)\lisvm$ & $\sim 10-20$  & $\simeq 120$ \\
 \hline
\end{tabular}
\caption{\footnotesize Well known resonances that affect lithium abundance. }
\end{center}
\end{table}

We now proceed with listing the important resonances that regulate
some of the reaction rates in (\ref{scaling}). We choose to list only
those resonances that have resonant energies one the order of $300$
keV or less, so that they are important at BBN energies. We organize
them in Table 1, where we follow the compilation of
Ref. \cite{A3-5} and \cite{A6-10}. The first reaction in this table
has an important contribution from the virtual level in the singlet
combination of $np$. It controls the deuterium bottleneck, the
temperature relevant for lithium production. The resonance in $^5$Li
controls the reaction $R_8$ which depletes \het, reducing its
availability for the $\alpha$-capture. Finally, the charge-exchange
reaction on \bes\ occurs right at the resonance in \beet. In addition
to those listed in Table \ref{t1}, there is also, of course, a
well-known resonance in $^5$He, $3/2^+$, with $E_X = 16.84$ MeV that dominates
the rate $R_7$ for the reaction $^3$H$(d,n)\hefm$ with resonant energy
being $\sim 50$ keV. It provides a secondary source of neutrons at
$T_9 \simeq 0.5 $.  It is also worth mentioning that the rate for
$R_6$, $\hetm(n, p)^3$H, is dominated by the sub-threshold resonance in
\hef. It is remarkable that any $O(1)$ variation in either resonant
energy or the widths of these resonances would translate to a similar
order change in the total
\bes\ abundance, Eq. (\ref{scaling}). Fortunately, at this point all
these reactions are very well known, both theoretically and
experimentally, and any $O(1)$ changes to their rates within the realm
of standard physics are simply not possible.

An interesting side remark is that a hypothetical possibility for reducing \het\ 
abundance via the resonant enhancement of $\hetm(\hetm,\alpha)pp$ reaction has been 
actively explored thirty years ago as a possible solution to the solar neutrino 
"under production" problem \cite{Be6th}. No additional resonances in $^6$Be 
and consequently no additional depletion of \het\ were found \cite{Be6exp}, 
and this conclusion can be directly carried over to the BBN calculation. 
In general, now well-measured neutrino flux summed over flavors agrees with the 
calculated abundances of \het\ in the Sun, which gives an indirect support to the BBN 
calculations of \het\ abundance.

We now proceed to searching for additional resonances in the secondary
reactions, not included in \cite{Cyburt}. These are the reactions of
direct burning of \bes\ by light elements other than $n$, such as
$p,~d,~t,~ \hetm$ and $\alpha$, which therefore should involve
resonances in such elements as boron and carbon. The resulting
possibilities for resonances found in \cite{A6-10} and \cite{A11-12}
are listed in Table \ref{t2}. Among these resonances, it is
immediately clear that the sub-threshold resonance in $^{11}$C cannot
play any role in the depletion of \bes\ because it is way too narrow,
$\Gamma/T \la 10^{-7}$. The other two cases cannot be immediately
discarded, as there is not enough experimental information about the
properties of these resonances. The 16.7 MeV resonance in \bor\ does
appear, however, as a more substantiated hope for reducing \bes\ then 
resonances in $^{10}$B, as
the abundance of deuterium at relevant temperatures is much larger
than that of tritium, $n_t /n_d \sim 10^{-2}$.

\begin{table}
\label{t2}
\begin{center}
\begin{tabular}{c|c|c|c}
Resonance & Reactions & $E_r$[keV] & $\Gamma_{cm}$[keV]\\
\hline\hline
 $\borm, ~{5/2^+}$, $16.7\pm 0.1$ MeV   & 
$\besm(d,\gamma)\borm$, $\besm(d,p)\alpha\alpha$  & $\sim 200$  & $\sim 40$  \\
 $^{10}$B, $2^+$, $18.8$ MeV& $\besm(t,\gamma)^{10}{\rm B}$, $\besm(t,p)^{9}{\rm Be}$,
$\besm(t,\hetm)\lisvm$ & $ \sim 130$  & $<600$ \\
 $^{11}$C, $3/2^+$, $7.50$ MeV &   $ \besm(\alpha,\gamma)^{11}{\rm C}$ & $-43$ & $< 10^{-4}$ \\
 
\hline
\end{tabular}
\caption{\footnotesize Resonances in boron and carbon that could potentially affect lithium abundance. }
\end{center}
\end{table}

\section{16.7 MeV 5/2$^{+}$ resonance in \bor\ and deuteron-induced 
reduction of \bes\ abundance}

Although the direct experimental information about properties of this
resonance is not available, something can be learned from the mirror
nucleus, \ben. There, the $5/2^+$ resonance at $16.671\pm 8$ MeV
energy is observed to be extremely narrow, $\Gamma = 41 \pm 4$ keV
\cite{Dixit}, presumably composed from $n$, $\alpha$ and $\gamma$
decay widths. It is reasonable to expect that a mirror resonance in
\bor\ is also narrow. In fact, $40$ keV is exactly the width that
separate "wide" from "narrow" for the BBN reactions, as the
temperature is also about the same value. The possibility of having a
resonance with $\Gamma \la T$ is important, as it leads to significant
variation of the astrophysical $S(E)$ in the relevant energy range. It
is worth mentioning that Ref. \cite{Angulo} made an explicit assumption
that $S(E)$ is energy independent below $E\sim 400$ keV. Since we know
that the position of this resonance is $200\pm 100$ keV above the
$\besm+d$ continuum threshold, one could expect significant variation
of $S(E)$ in the relevant energy domain, which in turn may compromise 
the extrapolation of the measurement \cite{Angulo} to lower energies.

In the absence of direct experimental information, at this point the
best strategy is to parameterize the properties of the resonance by
some values of the resonant energy, deuterium separation width and
total width $(E_r,~\Gamma_d,~\Gamma_{tot})$. We shall assume that
$\Gamma_{tot} \la 40$ keV so that approximation of the reaction rate
by a narrow resonance is appropriate. The statistical spin factor for
this reaction is $(2\times \fr{5}{2}+1)/(3\times 4)=1/2$, and the
total rate is given by
\ba
\label{rate}
R_{\besm+d} &=& \exp{\left( 26.429 - 12.5378/T_9^{1/3} - 6.86486T_9^{1/3} - T_9 -6.T_9^{5/3}\right)}/T_9^{2/3} \\ \nonumber
 &+& \exp{\left( 40.2944 - 21.1934/T_9^{1/3} - 5.781T_9^{1/3} + 0.14777T_9 - 0.00236773T_9^{5/3}\right)}/T_9^{2/3}\\ \nonumber
&+& \fr{3.9\times 10^7}{T_9^{3/2}} \times \fr{\Gamma}{1~{\rm keV}} \times
\exp\left(-\fr{2.32}{T_9}~\fr{E_r}{200~{\rm keV}}\right),
\ea 
where the rate is expressed in units of ${\rm s^{-1}{cm}^{3}/mole}$,
and $\Gamma \equiv \Gamma_d\Gamma_{out}/\Gamma_{tot}$. Notice that no
matter what the actual final state is (other than $d$), the \bes\
nucleus gets destroyed, and therefore we can add different pieces of
$\Gamma_{out}$ together, which for the Breit-Wigner resonance sums to
$\Gamma_{tot} - \Gamma_{d}$. To be consistent with the assumption of
narrow width approximation, $\Gamma$ has to be $\leq 40$ keV. The
non-resonant piece of $\besm(d,p)\alpha\alpha$ is updated according to
\cite{kavanagh60,Angulo}.

\begin{figure}
\centerline{\includegraphics[width=0.7\textwidth]{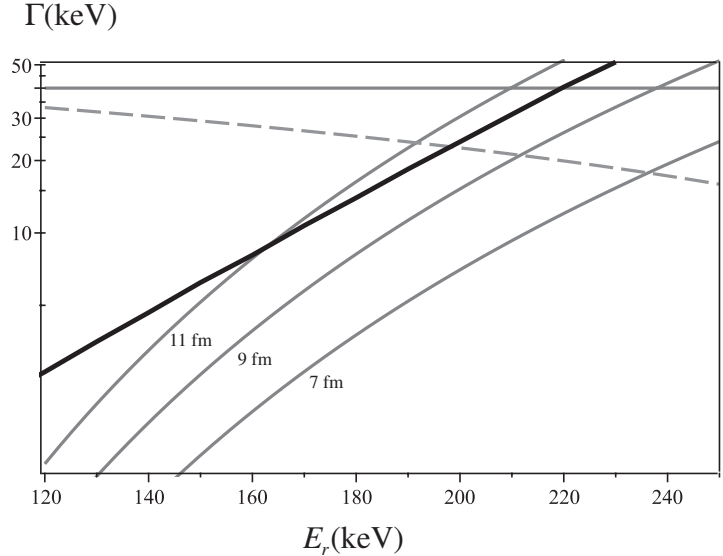}}
\caption{ Plotted are the resonance width, $\Gamma$ in keV vs the resonance 
energy, $E_r$ in keV.  The thick black line corresponds to
$\Gamma(E_r)$ at and above which the lithium problem is solved
(\lisv/H $\le 0.5\times [\lisvm/$H$]_{\rm SBBN}$).  Three solid diagonal lines are
the Wigner limits of $\Gamma$ for $a_{27} =7,9,11$ fm. The dashed line displays the 
sensitivity from \cite{Angulo} measurement (with the caveats pointed out in the text).}
\label{fig1}
\end{figure}

Plugging this rate into the full standard BBN code, we generate an
output for \bes+\lisv\ as a function of $E_r$ and $\Gamma$. In
Fig. \ref{fig1}, we invert this calculation and show
the curve of the constant depletion factor of 2 relative to the prediction
(\ref{BBNvalue}). One can easily see that on this plot there are
regions of parameter space that lead to the factor of 2 depletion of
\bes+\lisv\ while $\Gamma$ remains smaller than 40 keV, which would
correspond to the solution of the lithium problem by the resonant
enhancement of $\besm+d$ burning. At $E_r \simeq 220$ keV, the
required width $\Gamma$ becomes comparable to the temperature, where
our treatment of the narrow resonance is no longer valid. Therefore,
we consider 220 keV as an upper value of $E_r$ where the resonant
enhancement of \bes\ burning is capable of solving lithium problem.

Of course the most important question is whether $\Gamma$ can reach
the level required for the solution of lithium problem. This depends
on whether Coulomb suppression of the deuteron separation width
$\Gamma_d$ would preclude $\Gamma \sim O(10)$keV. The information on
mirror nucleus, \ben, is of no use in that respect, because \lisv+$d$
continuum threshold is above the 16.7 MeV $5/2^+$ resonance.
Therefore, inevitably one has to invoke some theoretical
considerations. As usual we define $\Gamma_d$ as a product of the
Coulomb penetration factor and reduced width $\gamma_d^2$,
\be
\label{Gammad}
\Gamma_d(E_r) = 2 \gamma^2_d P_1(E_r,a_{27}) \la  2 \gamma^2_{dW} P_1(E_r,a_{27}) \equiv \Gamma_{dW}.
\ee
In this expression, $P_1(E_r,a_{27})$ is the deuteron-\bes\ $p$-wave
Coulomb penetration factor, $a_{27}$ is the effective radius of the
entrance channel, and $\gamma^2_{dW}$ is the limiting Wigner
expression for the reduced width,
\be
\gamma^2_{dW} = \fr{3}{2\mu a_{27}^2},
\ee
and $\mu = m_{\besm}m_d/(m_{\besm} + m_d)$. 

Both the Coulomb penetration factor $P_1$ and the Wigner limit depend
quite sensitively on the channel radius $a_{27}$, that we keep as a
free parameter. Varying $a_{27}$, we plot the resulting {\em maximal}
$\Gamma_d^{max}$ on the same plot, as $\Gamma<\Gamma_d^{max}$. One can
immediately see that in order to be close to the required strength
$\Gamma$ capable of solving lithium problem, one has to go to the
unorthodox values of $a_{27}$ comparable to 10 fm. Nevertheless,
Fig. \ref{fig1} shows that {\em if} $a_{27} \ge 10$ fm, the strength
the quantum mechanically allowed value of $\Gamma$ can be above the
lithium problem solution line. This may happen only in the narrow
limit of energies, $180 \la E_r \la 220$ keV, and $\Gamma$ will have
to be comparable to 10 keV or larger values.

What are the reasonable values for the radius of the entrance channel? 
A default assumption for $a_{27}$ would be $a_{27} =
(2^{1/3}+7^{1/3})\times 1.4$ fm $= 4.4$ fm. This is an unrealistically
small value, as both nuclei in question, \bes\ and $d$, are quite
large. Another benchmark value is $a_d + a_{\besm} \simeq 6 $
fm. However, previous experience with the deuteron widths of low-lying
resonances, in particular with the $16.87 $ MeV resonance in $^5$Li,
suggest much larger interaction radii for $\hetm+d$ system \cite{Rmatrix}. The
$R$-matrix fit to this resonance uses $a_{23}$ to be 5 and 7 fm,
finding that for 5 fm the actual $\gamma_d^2$ would have to be {\em
above} Wigner limit. Given that $a_{\besm} > a_{\hetm} $ it is then
reasonable to allow $a_{27}$ exceed 7 fm. There is however, an
ultimate quantum-mechanical limit on $a_{27}$, coming from the
assumption of $a^2_{27} \sim 1/(E_r \mu)$ scaling. Adopting this
scaling, one can find that $a_{27}$ could be comparable to 12 fm for
$E_r \simeq 200$ keV, and therefore it is not inconceivable that
$\Gamma$ could reach 10 keV benchmark.

We find that the very fact that there exists a possibility for the
nuclear physics solution to the lithium problem via $\besm+d$
reactions is quite remarkable. If we go to the other possibility,
$\besm+t$, listed in Table \ref{t2}, we shall discover that no matter
what the properties of the $2^+$ resonance in $^{10}$B are, it cannot
lead to an appreciable depletion of \bes. Indeed, since the tritium
nuclei are less abundant than deuterium by a factor of $\sim 100$ at
relevant temperatures, the required $\Gamma$ would have to be
unrealistically large, comparable to an MeV, and would violate our
basic assumption of being a narrow resonance. Therefore, we are forced
to discard $\besm+t$ possibility.

An important question to ask is whether the contemplated resonant
enhancement could have been missed in the recent experiment that
remeasured $\besm(d,p)\alpha\alpha$ reaction rate \cite{Angulo}. There
is an explicit assumption made in this work that $S(E)$ is the smooth
function of energy below 400 keV. This assumption is violated by the
16.7 MeV resonance. Also, Ref. \cite{Angulo} detected only the very
energetic protons in the final state, which for example would miss
$\borm^*\to \beetm^*+p$ decays to the 16.63 MeV, $2^+$ level in \beet\
with the emission of $O(300-400)$ keV protons. Ref. \cite{Angulo}
argues that such decays will be sub-dominant to the decays into the
lower lying states of \beet\ because of the Coulomb suppression. This
is a valid argument for the continuum but may not necessarily work for
the $\borm^*(16.7) \to \beetm^*(16.63~{\rm MeV})+p $ transition. Should the decay of 16.7 MeV state in \bor\ indeed
proceed to that level in \beet\, the limits from \cite{Angulo} would
simply not apply. There
are also other possible final states such as \lisx+\het\ and
$\borm+\gamma$. In figure \ref{fig1}, we include the dashed line
indicating possible sensitivity of \cite{Angulo} to the resonant part
of this reaction, but this line should not be treated as a strict
upper limit.

\begin{figure}
\centerline{\includegraphics[width=0.95\textwidth]{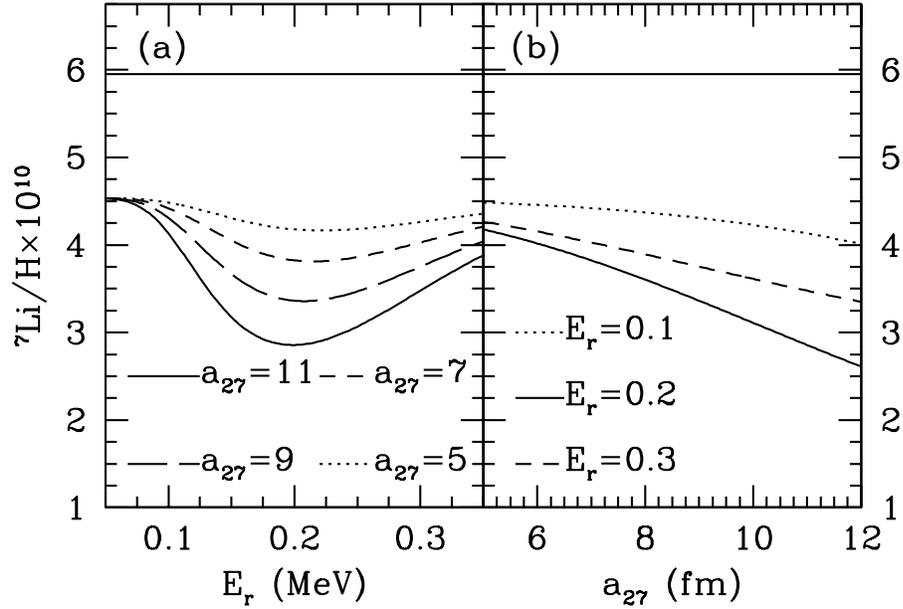}}
\caption{These figures show the standard BBN \lisv\ abundance predictions 
allowing for a resonance with energy $E_r$, with channel radius $a_{27}$.
Figure on the left shows the allowed 1-$\sigma$ range for fixed values of the
channel radius as a function of the resonance energy).  The channel
radius lies between 5 and $({200~ {\rm keV}\mu})^{-1/2}\sim 12$ fm .  Figure
on the right shows the allowed 1-$\sigma$ range for fixed resonance energies as
a function of the channel radius.}
\label{fig2}
\end{figure}

Apart from the question of whether or not the lithium problem is
solved by the resonant $\besm+d$ burning, it is important to quantify
the error bars in the \lisv+\bes\ prediction, as functions of the
parameters of the resonance. In what follows, we vary $100\la E_r \la
300 $ keV, and $ 0 \leq \Gamma \leq \Gamma_{dW}(a_{27})$, and
determine the range for the \lisv+\bes\ prediction as a function of
$a_{27}$. The results of this procedure are shown in Figure
\ref{fig2}. As one can see, the 1$\sigma$ band gets significantly
increased at $a_{27} \sim 9$ fm, and for $a_{27} \sim 12$ fm, a rather
wide range of answers is possible,
\be
\fr{\lisvm}{\rm H} = [2.5-6]\times 10^{-10}.
\ee
Therefore, assuming largest possible $a_{27}$ allowed by quantum mechanics, 
enlarges the uncertainty in predicting primordial lithium abundance 
by a factor of more than 2. 

\section{Conclusions}

Nuclear physics could in principle be responsible for the solution of the cosmological 
lithium problem. Narrow resonances may modify reactions between light elements and 
affect the abundances of \het\ and neutrons at $T\sim 40$ keV that directly affects 
the outcome of the \bes+\lisv\ synthesis. However, the reactions between elements
up to helium are well-known directly at BBN energies and the errors 
on the order of 100\% are implausible. It also appears fruitless to hypothesize
new previously unknown resonances in these reactions unless some significant experimental 
mistakes are made at $E_{cm} \sim 50$ keV. Therefore, the only chance of solving 
\bes\ overproduction is the direct destruction of this element by lighter species. 

If something destroys \bes\ in the early Universe, it is mostly likely
by an abundant species, namely neutrons, protons or
deuterons. Reactions involving neutrons are known way too well down to
very small energies and therefore cannot be a source of error in the
BBN calculation. Reaction involving protons, 
$\besm(p,\gamma)^8$B, is also very well known and its rate cannot catch
up with the Hubble rate at $T_9 \simeq 0.5$. The only remaining option
are deuterons (tritons are too rare).

In this paper, we have shown that the reaction rates of the \bes\
destruction by deuterons could be large, owing to a narrow resonance
$16.7$ keV, $5/2^+$ in \bor\ compound nucleus. This resonance may be
very strong, and at the very limit of the quantum mechanically allowed
value for the deuteron separation width, which would be responsible
for a factor of $\sim 2$ suppression of the primordial \bes\ yield,
thus resolving the lithium problem. This resonance is presumably
somewhat below the range of energies probed by experiment
\cite{Angulo}, and due to a non-monotonic dependence of $S$-factor on
energy, could have been missed.

If indeed lithium problem is resolved this way, it means that the
actual size of $5/2^+$ \bor\ nucleus is very large, as large as 12 fm,
and to a large extent this state should be represented by the $p$-wave
bound state of deuteron and \bes. The width of this state, possibly as
narrow as 40 keV, should have large contributions from $\Gamma_d$, 
and the rest come from $\Gamma_\gamma, \Gamma_p$ and $\Gamma_\alpha$.

In the absence of direct experimental information about this resonance
level, and in the absence of dedicated theoretical nuclear studies of
its properties, it is of course impossible to declare that this is the
solution to the lithium problem. At the same time our study shows that
it is premature to rule out the nuclear physics solution to \lisv\
problem. Only the dedicated study of the $5/2^+$
state in \bor\ can resolve this issue. Being completely agnostic about
the size of the entrance channels, $a_{27}$, and varying the
parameters of the resonance constrained only by quantum mechanics, we
find that the error bars in the prediction of lithium abundance should
be enlarged, and the whole range $(2.5-6)\times 10^{-10}$ is possible.

Finally, we comment on the possibility of experimental test of 
our hypothesis. 
There are several experimental possibilities of activating the 16.7 MeV
state in \bor.  Inelastic scattering experiments, such as
($e,e^{\prime}$) and ($p,p^{\prime}$) can be used to populate states
in the mirror nuclide \ben.  Employing isospin symmetry one can gain
further insights into the physics of \bor\ nuclide. One can also use 
charge exchange reactions on \ben\ to populate states
in \bor, such as the reactions ($p,n$) and ($\hetm,t$).  Stripping
reactions may also be used to populate this state, via the
($p,\alpha$), ($\hetm,\lisxm$) and ($\alpha,\lisvm$) reactions on the
\ctw\ nucleus.  Key to these studies would be the use of $\gamma$ and
particle coincidences to tag states.

\subsection*{Acknowledgements}
RC would like to thank S.~Austin, B.D.~Fields, K.A.~Olive, H.~Schatz
and A.W.~Steiner for useful discussions.  The work of RC was
supported by the U.S. National Science Foundation Grants
No. PHY-01-10253 (NSCL) and No. PHY-02-016783 (JINA).  The work of MP
was supported in part by NSERC, Canada. Research at the Perimeter
Institute is also supported in part by the Government of Canada
through NSERC and by the Province of Ontario through MEDT.

\end{document}